\documentclass[]{aa}
\usepackage{amsmath}
\usepackage{graphicx}
\usepackage{natbib}
\usepackage{xcolor}
\bibpunct{(}{)}{;}{a}{}{,}
\usepackage[american]{babel}
\usepackage{txfonts}

\begin{document}

\title{Formation of phase lags at the cyclotron energies in the pulse
  profiles of magnetized, accreting neutron stars}

\author{G. Sch\"onherr\inst{1,2}
        \and
        F.-W. Schwarm\inst{2}
        \and
        S. Falkner
        \inst{2}
        \and
        T. Dauser
        \inst{2}
        \and
        C. Ferrigno
        \inst{3}
        \and
        M. K\"uhnel
        \inst{2}
        \and
        D. Klochkov
        \inst{4}
        \and
        P. Kretschmar
        \inst{5}
        \and
        P.A.\ Becker
        \inst{6}
        \and
        M.T.\ Wolff
        \inst{7}
        \and
        K. Pottschmidt\inst{8}
       \and
        M. Falanga\inst{9}
        \and
        I. Kreykenbohm\inst{2}
        \and
        F. F\"urst\inst{10}
       \and
       R. Staubert\inst{4}
        \and
        J. Wilms\inst{2}
      }

\authorrunning{G. Sch\"onherr et al.}
\titlerunning{Formation of phase lags at cyclotron energies}
  \offprints{G. Sch\"onherr}
\institute{
Leibniz-Institut f\"ur Astrophysik Potsdam (AIP), An der Sternwarte 16, 14482 Potsdam, Germany, \email{g.schoenherr@aip.de}
        \and
Dr.~Remeis-Sternwarte \& ECAP, Sternwartstr.~7, 96049 Bamberg, Germany
        \and
INTEGRAL Science Data Centre, Universit\'e de Gen\`eve, Chemin d'\'Ecogia 16, 1290
Versoix, Switzerland 
        \and
Institut f\"ur Astronomie und Astrophysik, Abt.\ Astronomie, Universit\"at T\"ubingen,
Sand 1, 72076 T\"ubingen, Germany
        \and
European Space Astronomy Centre (ESA/ESAC), Science Operations
Dept., P.O.\ Box 78, 28691 Villanueva de la Ca\~nada, Madrid,
Spain 
        \and
George Mason University, 4400 University Drive, Fairfax, VA 22030, USA
        \and
High Energy Space Environment Branch, Space Science Division, Naval
Research Laboratory, Washington DC 20375, USA
        \and
CRESST, University of Maryland Baltimore County and NASA's Goddard
Space Flight Center, Greenbelt, MD 20771, USA 
\and International Space Science Institute (ISSI),
Hallerstr.\ 6, 3012 Bern, Switzerland
\and California Institute of Technology, Pasadena, CA 91125, USA
}

\date{Received ---; accepted ---}

\abstract{Accretion-powered X-ray pulsars show highly energy-dependent
  and complex pulse-profile morphologies. Significant deviations from
  the average pulse profile can appear, in particular close to the
  cyclotron line energies. These deviations can be described as energy-dependent phase lags, that is, as energy-dependent shifts of main
  features in the pulse profile.}{Using a numerical study we explore
  the effect of cyclotron resonant scattering on observable,
  energy-resolved pulse profiles. }{We generated the observable
  emission as a function of spin phase, using Monte Carlo
  simulations for cyclotron resonant scattering and a numerical ray-tracing routine accounting for general relativistic light-bending
  effects on the intrinsic emission from the accretion columns. }{We
  find strong changes in the pulse profile coincident with the
  cyclotron line energies. Features in the pulse profile vary strongly
  with respect to the average pulse profile with the observing
  geometry and shift and smear out in energy additionally when assuming a
  non-static plasma.}{We demonstrate how phase lags at the cyclotron
  energies arise as a consequence of the effects of angular
  redistribution of X-rays by cyclotron resonance scattering in a
  strong magnetic field combined with relativistic effects. We also
  show that phase lags are strongly dependent on the accretion
  geometry. These intrinsic effects will in principle allow us to
  constrain a system's accretion geometry.}


\maketitle

\section{Introduction}
\label{sec:intro}

Accreting X-ray pulsars \citep[see, e.g.,][for a recent
review]{caballero:12a} exhibit complex and highly energy-dependent
pulse profiles \citep{lutovinov:09a}. These profiles can vary with
time and luminosity, for instance over outbursts of transient sources
\citep[e.g.,][]{tsygankov:06a} or in the case of Her X-1 with a
super-orbital period \citep{staubert:13a}. However, the detailed
physical origin for this complexity is still poorly understood. A
number of efforts in the past have attempted to disentangle the components of intrinsic
beam patterns for individudal sources by a complex ``backward''
approach as described by \citet{kraus:95a}, by Fourier-decomposing the
observed pulse profiles into individual contributions of radiation
from the poles \citep[e.g.,][]{kraus:96a, caballero:11a, sasaki:12a}.
While this approach has proven to be very successful in offering
possible solutions for the beam patterns, that is, the angle- and
energy-dependent emissivity of X-rays at each pole, it does not
address the physical processes that yield these beam patterns, nor does
it result in a unique choice among several possible solutions.

The most striking feature of many observations is that X-ray pulse
profiles are strongly energy dependent. This is especially apparent in
the energy-dependent location of the main pulse peak, which changes
significantly in pulse phase, that is, the main peak shows a phase lag
\citep{tsygankov:07a,tsygankov:06a,ferrigno:11a}. This phase lag, relative to the mean pulse profile, is
most apparent around the energy of the cyclotron line, at
energies where photons can scatter resonantly with electrons on Landau
orbits in the strong magnetic field of the accretion column and where
the scattering cross section is strongly angle dependent.

In this \textsl{Letter}, we present a new approach in which pulse
profiles are calculated based on a self-consistent physical model of
the intrinsic radiation pattern at the two magnetic poles, including
all general relativistic effects such as light bending or
gravitational redshifting (Sect.~\ref{model}). This approach follows
early fundamental works on the radiation from cyclotron line sources
\citep[see, e.g.,][and references therein]{meszaros:88b,sturner:94a,zheleznyakov:86a, soffel:85a}, but differs
in numerical method. This
allows for the resolution of model spectra and pulse profiles at a new level, as we
illustrate here for one characteristic observable, energy-dependent
pulse profiles (Sect.~\ref{pp_maps}). We explore under which
conditions intrinsic deviations from the mean, energy averaged pulse profile due to cyclotron resonance scattering lead to the formation of
characteristic phase lags relative to the mean pulse
profile at the cyclotron energies.

\section{Model setup}
\label{model}
 
\begin{figure}
  \centering
  \resizebox{0.8\hsize}{!}{\includegraphics{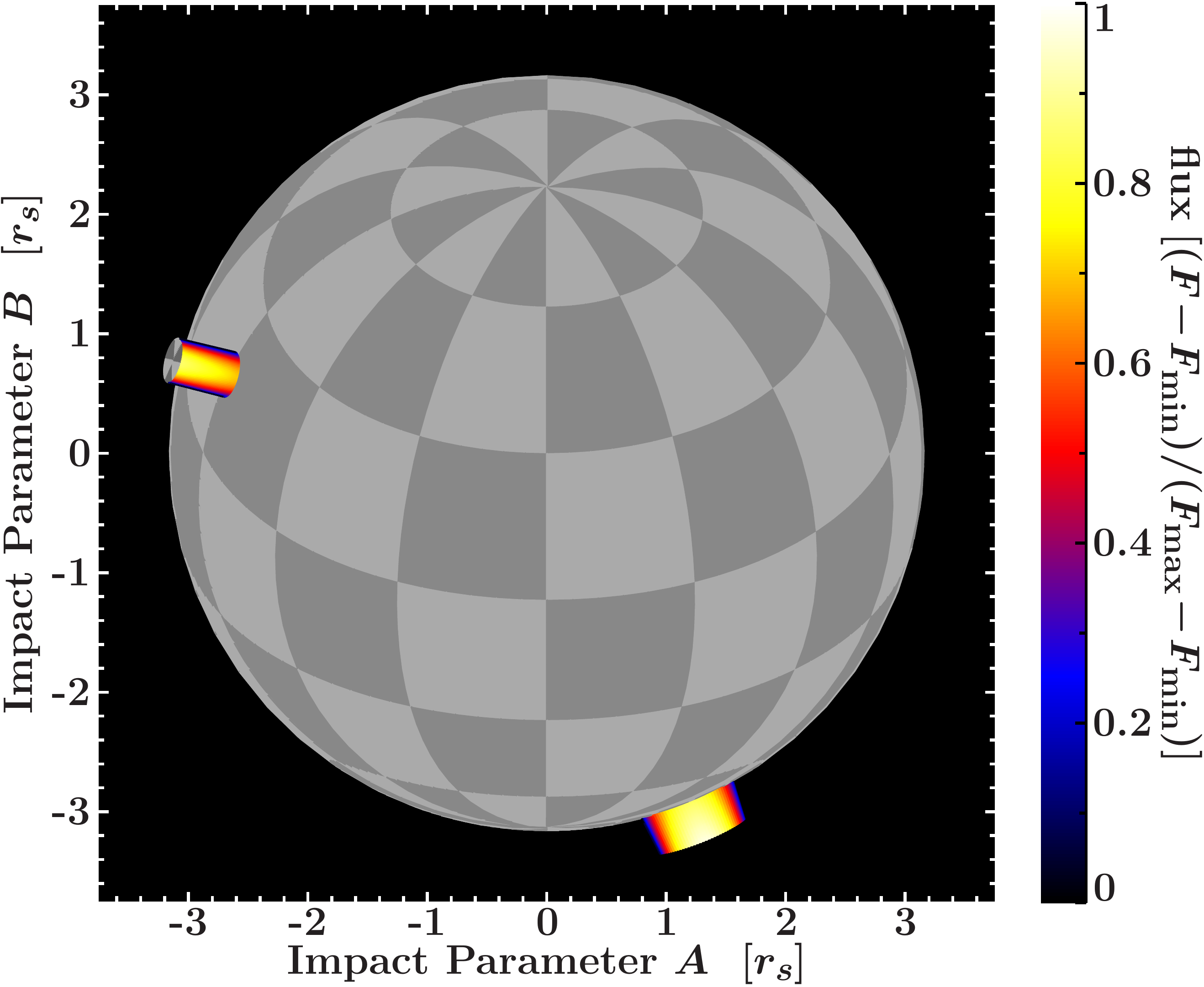}}
  \caption{Observer's view of a neutron star, illustrating the general
    relativistic effects on the visibility and flux of two accretion
    columns for an observer at infinite distance from the neutron
    star. Sky coordinates are parameterized through the impact
    parameters $A$ and $B$, which correspond to the projected distance
    from the gravitational center in units of the Schwarzschild
    radius, $r_\mathrm{s}$. The illustration corresponds to the
      two-pole geometry used in Sect.~\ref{sect:asym}. Different
    apparent sizes of the accretion columns are due to strong light-bending effects.}
  \label{fig:bp}
\end{figure}

We considered a canonical neutron star ($M=1.4\,M_{\odot}$, $R=10$\,km)
with a strong magnetic field on the order of several $10^{12}$\,G
that accretes matter onto two magnetic poles (see Fig.~\ref{fig:bp}). We furthermore assumed that we observe only lateral emission from the walls of possibly elongated accretion columns, a ``fan beam scenario''. For sources with clearly defined cyclotron lines, this necessarily induces the postulation of a constrained 
``efficient area of the accretion column to line formation'' \citep{nishimura:08a}, which must be of practicable negligible magnetic field gradient.
The X-ray emitting region of the accretion column at each pole is
 therefore modeled by a simple, homogeneous, small cylindrical volume. The location, radius, and
height of the cylindrical volume, the location of the accretion columns
with respect to the spin axis of the neutron star, and the inclination
of the system with respect to the observer are free model parameters. We do not address any possible ``pencil beam'' radiation components, nor do we consider surface reflection \citep[see, e.g.,][]{poutanen:13a} for this pilot study.

The plasma electrons have a Maxwellian temperature distribution
parallel to the $B$-field and are magnetically quantized in their
perpendicular momenta. We generated the intrinsic beam pattern at each
pole by reprocessing X-ray continuum seed photons from the column by
cyclotron resonant scattering between electrons and X-ray photons in
its outer layer. This picture is motivated by the comparably very
small mean free path of resonant photons that translates into a thin
``scattering atmosphere'' around the optically thick core. The
scattering of photons and electrons causes cyclotron resonant
scattering features (if observable) in the source spectrum at $E_{n,
  \mathrm{cyc}}\sim n \, 11.6\,B\,{[10^{12}}\,$G$]/
(1+z)\,\mathrm{keV}$, where $z$ is the gravitational redshift and $n$
is an integer. In the following we assume $B= 4 \times
10^{12}$\,G, a parallel electron temperature of
$kT_\mathrm{e}=3$\,keV, and a cyclotron scattering atmosphere with
Thomson optical depth $\tau =10^{-3}$ perpendicular to the $B$-field
axis \citep[see, e.g.,][]{schoenherr:07a, suchy:08a}.

\begin{figure*}\sidecaption
        \centering
        \includegraphics[width=5.5cm]{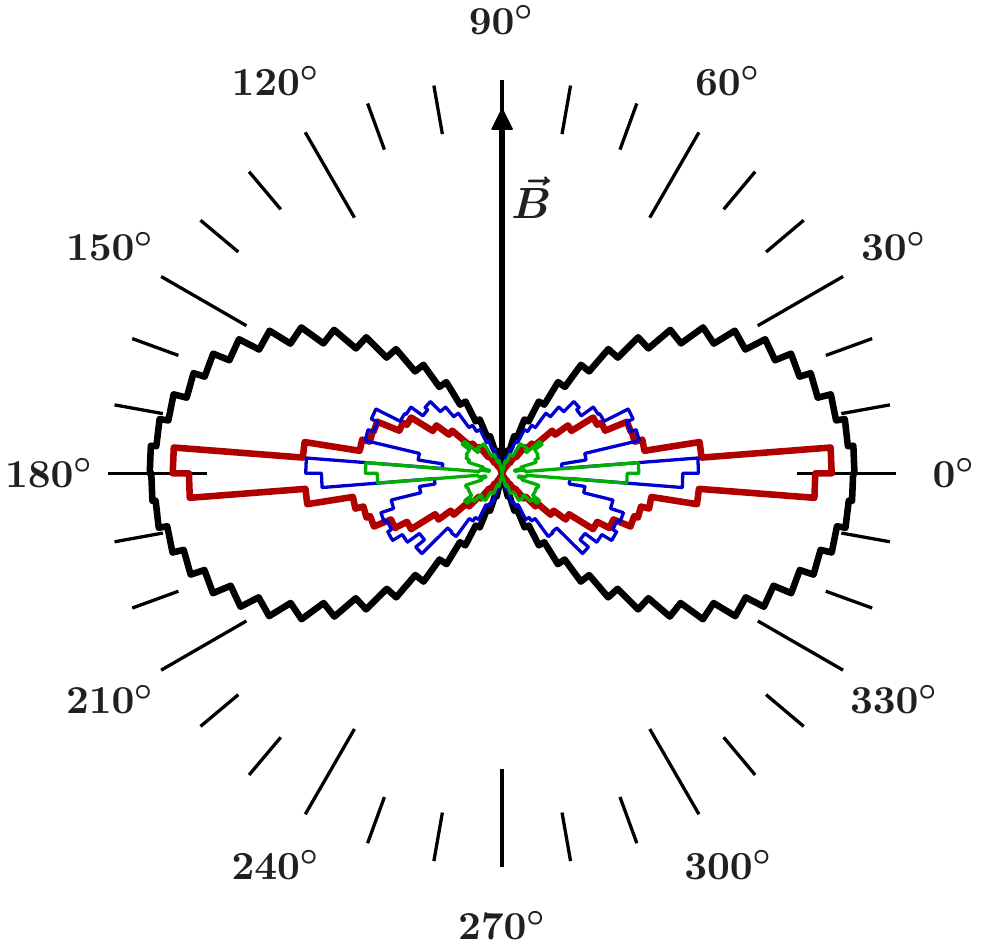}
        \includegraphics[width=5.5cm]{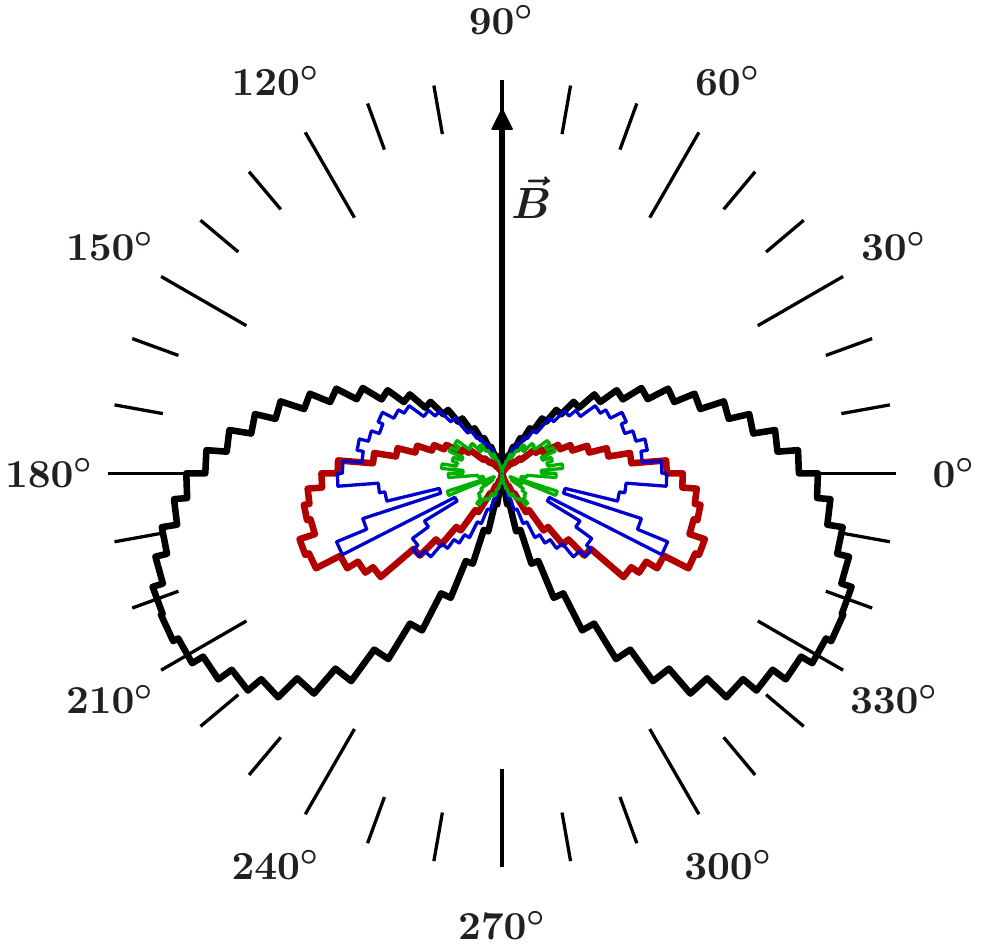}
        \caption{ \label{fig:bp0} Energy-resolved beam patterns after
          cyclotron scattering for a cylindrical geometry (``fan
          beam''). \textbf{Left:} case of a static plasma.
          \textbf{Right:} plasma with bulk velocity $v=0.4c$. The
          black line shows the angular distribution of all photons.
          The red, green, and blue lines mark photons escaping at
          energies close to the cyclotron resonances for the
            respective angles (not to scale). While the photons
            that escape at the fundamental resonance (red) have all
            undergone scattering, the photons escaping at the higher
            resonances are unaltered from the continuum
            \citep[see][]{schwarm:13a}.} 
\end{figure*} 

Because we wish to investigate the isolated effects of beaming by cyclotron
resonant scattering on the observable pulse profiles separated from
possible continuum effects, we assumed the continuum radiation to be
isotropic and to have a simple power-law spectrum with a photon index
$\Gamma = 1$. The assumption of isotropy is justified as the
  intrinsic beaming of the continuum radiation is expected to be
comparably weak. For a fast falling plasma, an overall downward
  beaming has to be taken into account as well
  \citep[see][]{lyubarsky:88a}. The energy and angular redistribution
of the continuum while passing the cyclotron scattering layer were
calculated based on Monte Carlo simulations of
\citet[][2013]{schwarm:10a}, which follow the same fundamental
  physical setup as described in detail by \citet[][]{araya:99a}, \citet[][]{schoenherr:07a}, and references therein. Fully relativistic cross-sections \citep{sina:96a} were used to describe the
  polarization-averaged scattering process. Again, to bring out the
effects of cyclotron scattering on the pulse profiles, identical
optical depths and $B$-field strengths were assumed for both poles. We
obtain the flux emitted from each point of the surface of the walls of
the cylindrical volume as a function of energy and angle with respect
to the $B$-field.

We considered both static volumes and a falling plasma with a bulk
velocity of $0.4c$ (see Fig.~\ref{fig:bp0}). The effect of a bulk
velocity on our simulation results is discussed in more detail
by \citet{schwarm:13a}. For earlier work on cyclotron
  resonant scattering in non-static plasmas see, e.g.,
\citet{soffel:85a}, \citet{isenberg:98a}, and \citet{serber:00a}.
As the large-scale $B$-field structure of High-Mass X-ray Binaries and the magnetospheric
coupling of the accretion flow are poorly constrained, we considered
both a symmetric geometry and a geometry where the magnetic poles are
displaced in longitude and latitude with respect to an antipodal
setting. The latter has been shown to be one option to generate
asymmetric pulse profiles that resemble observed ones from otherwise
symmetric assumptions \citep[e.g.,][]{bulik:95a, kraus:96a,
  caballero:11a, sasaki:12a}. We calculated the observed time and
energy-dependent flux from these setups with a fully relativistic
ray-tracing code that includes light bending, gravitational
redshifting, and special relativistic effects due to the motion of the
emitter \citep{falkner:13b}.

\section{Energy-dependent pulse profiles}\label{pp_maps}

\subsection{Symmetric geometry}
\label{sect:sym}

\begin{figure*}\sidecaption
  \centering
  \includegraphics[width=12.5cm]{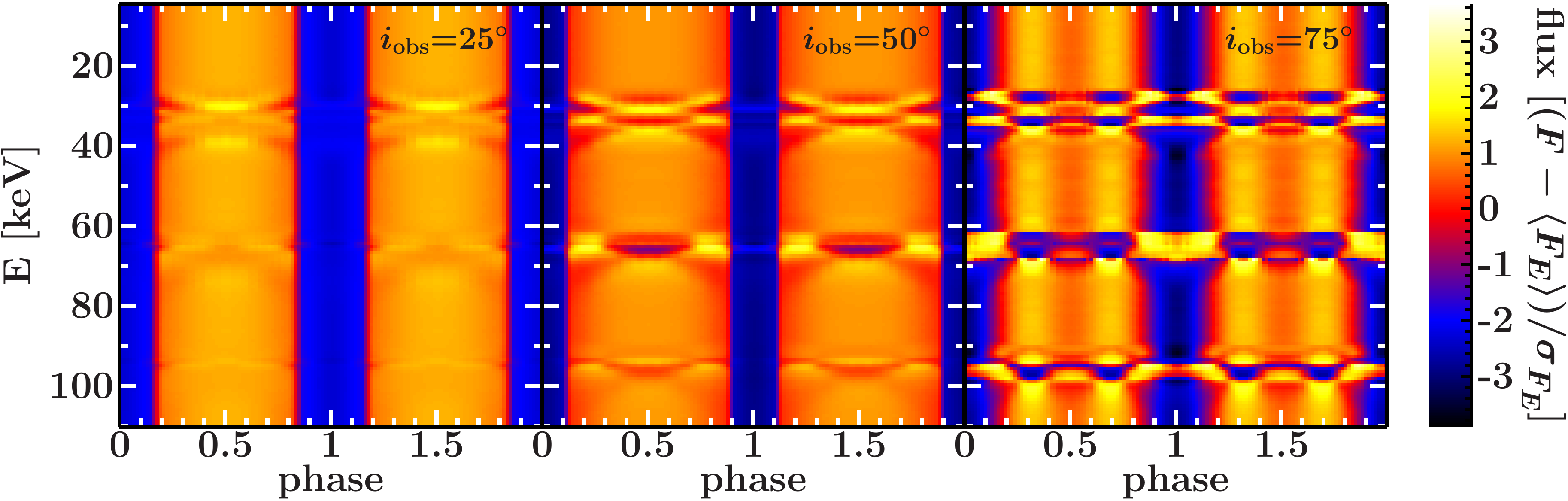}
  \caption{\label{fig:maps_obs}Energy-dependent flux as a
    function of pulse phase for different observation angles
    $i_\mathrm{obs}=25^\circ$, $50^\circ$, and $75^\circ$ from left to
    right (see Sect.~\ref{sect:sym} for the geometrical setup). The pulse
    profiles are normalized such that for every energy bin the mean flux
    is zero and the corresponding standard
    deviation is unity \citep[after][]{ferrigno:11a}.
    The pulse profiles are plotted twice for clarity.}
\end{figure*}

First, we discuss a symmetric, antipodal geometry of two static
accretion columns of radius $r_1=r_2=1$\,km and height
$h_1=h_2=100$\,m, which are displaced at an angle $i_1 = i_2 =
35^\circ$ with respect to the spin axis of the neutron star. Because
  we assumed a constant magnetic $B$-field, the heights $h_i$ do not
  represent a realistic accretion column structure. They instead rather
  reflect the uncertainty of the height from which emission from a
  sufficiently constrained line-forming region is observed, which for
  this example was assumed to be somewhere at the base of the column
  not farther than $100\,$m from the surface. 
Figure~\ref{fig:maps_obs} shows the resulting energy-dependent pulse
profiles for different observing angles. Horizontal cuts through the
figures represent pulse profiles at a given energy, while vertical
cuts represent the deviation of the X-ray spectrum from a normalizing
continuum at a given pulse phase. Observing this system at a line of
sight tilted by $i_\mathrm{obs}=25^\circ$ with respect to the
rotational axis results in practically no features being apparent from which one might distinguish between the relative pulse profiles at different
energies. Observations would show one symmetric broad pulse over the
entire energy range. With increasing $i_\mathrm{obs}$, the main peak
of the pulse profile becomes narrower and strong deviations of the
pulse profiles appear around the redshifted cyclotron line energies
($\sim$32, 64, and 96\,keV). These changes can be understood from the
very strong perpendicular beaming of photons that escape at the
cyclotron energies (Fig.~\ref{fig:bp0}). These photons can come into
view of the observer, for a suitable geometric setup, at a pulse phase
when most of the main beam at all other energies is still hidden.

\subsection{``Wavy'' phase lags}
\label{sect:asym}

Motivated by an observational study of phase lags, we demonstrate the
effects of strong asymmetry of the location of the accretion columns
and of a non-static plasma for one specific example: To explain the
strong changes in the highly asymmetric and variable pulse profiles of
the accreting X-ray pulsar \object{4U0115+63}, \citet{ferrigno:11a}
invoked a two-pole geometry with $i_1 =74^\circ$, $i_2=32^\circ$ and a
phase offset of $68^\circ$ for an observation angle of
$i_\mathrm{obs}=60^\circ$, as was proposed by
\citet{sasaki:12a} from applying the pulse decomposition method to
this source. The authors additionally assumed laterally emitting
columns of $r_1=r_2 \sim 700$\,m and $h_1=h_2=2$\,km
(Fig.~\ref{fig:bp}) and heuristically modeled the intrinsic emissivity
pattern from these columns as the sum of an upward and a downward
pointing beam with a Gaussian emissivity profile. From purely
geometric considerations they proposed that a relatively enhanced
upward component at the cyclotron line energies might be causing the
observed pulse profile changes and the associated strong phase shifts.

\begin{figure*}\sidecaption
  \centering
  \includegraphics[width=12.5cm]{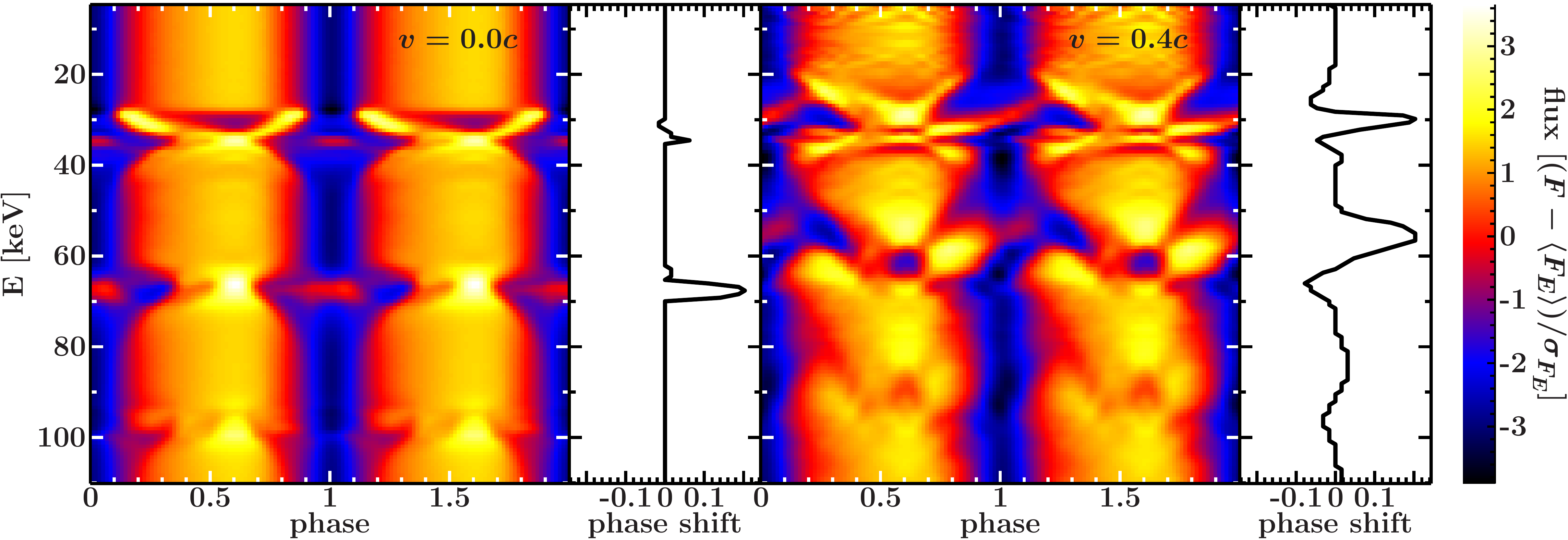}
  \caption{Color plots: flux maps for the geometry of
    Fig.~\ref{fig:bp}. The additional white panels show the
    corresponding energy-dependent phase shifts as calculated with
    respect to the full energy pulse profile using cross-correlation
    as described in detail and applied to observational data by
    \citet{ferrigno:11a}. \textbf{Left:} Static plasma.
    \textbf{Right:} Downward bulk velocity of $v=0.4c$. The values of
    $B$, $kT_\mathrm{e}$, and $\tau$ as well as the flux scaling are
    the same as in Fig.~\ref{fig:maps_obs} (see
    Sect.~\ref{sect:sym}).}
\label{fig:wave}
\end{figure*}

We now adopt the same parameters for the two-pole accretion column
geometry of our numerical model. The relative fluxes generated
  from a constrained emitting volume of given parameters -- in
  particular a given value for the magnetic field $B$ for a source
  with sharp cyclotron lines -- located at arbitrary height between
  the top of the accretion mound and $2\,$km up in the column above
  the surface show very little variation. Therefore, we used spatially
  averaged fluxes (Fig.~\ref{fig:wave}) for our further analysis.
In contrast to \citet{ferrigno:11a} we did not assume any artificially
imposed beamed emission patterns. Instead, as discussed for the
symmetric setting in the previous paragraph, we used the simulated
intrinsic beam patterns of the photons that were redistributed
in angle by cyclotron resonant scattering \citep[see
Fig.~\ref{fig:bp0} and][]{schoenherr:07a,schwarm:13a}.
Figure~\ref{fig:wave} shows the energy-dependent pulse profiles and
associated phase lags for a static plasma and for a non-static plasma
with a bulk velocity of $v=0.4c$ within the line-forming region.
While for the static plasma the phase shifts of the peak at the
cyclotron energies relative to the mean pulse profile are very
localized and rather small, the introduction of a bulk velocity
component yields a broad and wavy morphology similar to the observed
one \citep[see Figs.~3 and 5 of][]{ferrigno:11a}.

A fast-falling plasma can significantly enhance the angular difference
in beaming between the contribution from photons at the cyclotron
fundamental resonance and the overall radiation: for a static
plasma the photons at the cyclotron fundamental mostly escape
perpendicular to the field (Fig.~\ref{fig:bp0}, left) and only the
neutron star surface introduces an asymmetry with respect to the
overall radiation. For the case of bulk velocity a broader downward
beam at the cyclotron fundamental and upward beams at the cyclotron
higher harmonics are generated (Fig.~\ref{fig:bp0}, right). Hence, the
phase dependence of peaks at the fundamental and the harmonic
cyclotron features are expected to differ for an appropriate geometry,
as has been seen in 4U0115+63 \citep{tsygankov:07a}. This might also
explain the observed different energy localisation of the maximum
negative phase displacement of the pulse profile peak with respect to
the fundamental and first harmonic scattering feature in the study by
Ferrigno et al. \citep[Fig.~7 and~8 of][]{ferrigno:11a}.

\section{Discussion and conclusions}

The fact that we still lack basic knowledge about the coupling of the
accretion flow to the magnetosphere, which determines the location and
geometry of the emitting regions, unfortunately adds a significant
amount of freedom to theoretical predictions of pulse profiles of
X-ray pulsars. It is therefore important to identify and understand
characteristic trends in the simulations that are comparable to
observable ones to narrow down the physically reasonable
parameter space and to prepare the grounds for future proper fits of a
model to observations. Focusing on the investigation of the
formation of strong changes in the pulse profile at the cyclotron
energies, we have shown how phase shifts relative to the energy-averaged pulse profile arise naturally from the intrinsic beaming by
cyclotron resonant scattering processes for suitable geometrical
setups. The assumption of a static versus a non-static plasma has
proven to be highly relevant for the overall pulse profile morphology.
A next step will be to investigate more complex geometries, for example height-dependent velocity gradients, 
temperature profiles and $B$-field geometries. Especially for the non-static case the relevance of a possible reflection component from the surface due to strong downward beaming of radiation needs to be investigated. The understanding of
scattering related, apparent phase lags allows us to separate their
physical origin from other geometric effects that can cause changes in
the morphology of the energy-dependent pulse profiles such as a
height-dependent continuum or occultations by the accretion flow.

\begin{acknowledgements}
  We thank the International Space Science Institute ISSI in Bern (CH)
  for granting two International Team meetings on ``The physics of the
  accretion column of X-ray pulsars'', which have much inspired this
  collaborative work. We also thank the Bundesministerium f\"ur
  Wirtschaft und Technologie for funding through Deutsches Zentrum
  f\"ur Luft- und Raumfahrt grant 50\,OR\,1113. MTW is supported by
  the US Office of Naval Research and the NASA ADAP Program under
  grant NNH13AV18I. We thank the anonymous referee for very
    useful comments.
\end{acknowledgements}

\end{document}